\documentstyle[aps,preprint]{revtex}
\begin{document}
\draft

\title{
Influence of hydrodynamic interactions on the adsorption process
of large particles.
}

\author{I. Pagonabarraga and J.M. Rub\'{\i}}

\address{
Departament de F\'{\i}sica Fonamental, Facultat de F\'{\i}sica \\
Universitat de Barcelona\\
Diagonal 647, 08028 Barcelona, Spain
}

\date{\today}

\maketitle

\begin{abstract} We have studied the adsorption process of
non-Brownian particles on a line.  As a new ingredient with respect
to
previously proposed models, we have incorporated hydrodynamic
interactions between the incoming particles and the preadsorbed
particles as well as the surface.  We then numerically analyze the
effect of these interactions on typical relevant quantities related
to
the adsorption process.  Comparing our model to the ballistic
deposition model (BM) shows in particular a significant discrepancy
in
the pair correlation function.  The results obtained can explain some
differences observed between recent experiments and BM predictions,
as
reported in ref.\cite{Sch1}.  Finally, the limitations of the
applicability of BM in experimental situations is addressed.
\end{abstract}


\narrowtext

	The adsorption of large particles on solid surfaces is a
problem which has been subject to a great deal of interest during the
last years.  This process is mainly constituted by two basic steps:
the transport of the objects from the bulk towards the surface, and
their subsequent adhesion to it.  A number of models have been
proposed in recent years in an attempt to describe the process.  In
particular, the random sequential adsorption model (RSA) was
initially
introduced as a simple model which captures the essentials of its
kinetics \cite{Hir}.  In RSA the transport of particles to the
surface
is not considered.  Thus, a particle arriving at the surface is
accepted if it does not overlap with a previously adsorbed one;
otherwise it is rejected.  Specific quantities, such as the maximum
coverage of the surface, or jamming limit, are in accordance with
some
experimental results \cite{Ono}\cite{Ram}\cite{Adam}, so it was
concluded that RSA was a good model when particles diffuse to the
surface.  However, further numerical studies which took into account
the diffusion showed discrepancies in the pair distribution
function\cite{Sen}.  To check the validity of RSA, experiments on the
deposition of colloidal particles have been performed\cite{Sch1}.  In
these experiments, gravity has been shown to play an important role,
and therefore it seems more appropriate to compare with the results
predicted by the ballistic model (BM)\cite{Tal}.  In this model,
particles approach the surface following straight trajectories, and
are accepted if no preadsorbed particle is present or if there is
room
at its adjacent region.  In the latter case, the particles roll down
the surface of the preadsorbed ones\cite{bm1}.  If there is no room,
the particle is rejected.  Comparison of the pair distribution
function observed experimentally with BM results exhibits some
discrepancies, which may, in principle, be attributed to
polydespersity, van der Waals forces (vdW) or hydrodynamic
interactions
(HI).  Preliminary results\cite{Sch2}, however, have shown that
polydispersity does not explain this disagreement.

	Our purpose in this Letter is to address the effects of HI in
the adsorption process.  Up to now, HI have only been considered in a
situation in which gravity can be neglected \cite{Bafa}.  In this
case, as the particle diffuses more easily paralelly rather than
perpendicularly to the plane due to HI, no strong differences are
observed with RSA predictions.  However, if gravity is present, the
randomization effect of the diffusion dissapears, and, therefore, no
guess can be made about the importance of HI on the adsorption
kinetics.

	To elucidate the effects of HI we have proposed a
one-dimensional model in which spherical particles of radius $a$
falling down one by one towards the adsorbing line due to the
presence
of a gravity field, interact with the nearest preadsorbed particles
and with the surface.  When they reach the line they are accepted on
the surface if there is enough room for them; otherwise, they are
rejected.  Once the particles are adsorbed, they stick on the
surface.
In this respect, the kinetics is quite similar to the one prescribed
by BM, although the rolling now can be performed without touching the
adsorbed sphere (see Fig.  1a).  Moreover, during the process the
incoming particle is assumed to interact only with its nearest
neighbors on the line.  We have analysed our system under the
conditions such that inertial and diffusion effects may be neglected.
Finally, in the treatment of HI we have assumed additivity of the
friction tensors.  This standard approximation, already introduced in
Ref.\cite{Bos} \cite{Bafa}, considerably simplifies our treatment,
and it is
sufficient to make evident the effects of HI in the adsorption
process.

        Under these approximations, the dynamics of the incoming
particle is governed by the equation

\begin{equation}
\vec{v}=\vec{\vec{\mu}}\cdot \vec{F}\equiv m_{app}\vec{\vec{\mu}}
\cdot \vec{g}
                                                \label{eq1}
\end{equation}

\noindent Here $\vec{v}$ is the velocity of the particle,
$\vec{\vec{\mu}}$ its mobility tensor, $\vec{g}$ the acceleration of
gravity, and $m_{app}=4 \pi a^{3} \Delta\rho/3$ the apparent mass,
with $\Delta\rho\equiv(\rho_{p}-\rho_{f})$, $\rho_{p}$ being the
density of the particle and $\rho_{f}$ the density of the fluid.
Note
that in eq.(\ref{eq1}) we have not considered van der Waals forces
with the plane since, contrary to what happens for Brownian
particles\cite{Bafa}, they are not relevant in the dynamics of large
particles due to its short range nature.  The expression of the
mobility matrix takes into account the existence of HI.  Under the
additivity approximation, the friction tensor $\vec{\vec{\xi}}$,
which
is the inverse of the mobility matrix, splits up into contributions
due to the surface and to the already attached spheres.  As only
nearest neighbor interactions are assumed, this last contribution is
reduced to the contributions coming from the presence of the two
nearest adsorbed spheres.  One has

\begin{equation}
\vec{\vec{\xi}}=\vec{\vec{\xi}}_{sp}+\vec{\vec{\xi}}_{s1}+
\vec{\vec{\xi}}_{s2}-2 \xi \vec{\vec{1}}
                                                \label{xi1}
\end{equation}

\noindent where $\vec{\vec{\xi}}_{sp}$ is the friction tensor of one
sphere in the presence of a plane, without any other sphere being
present, and $\vec{\vec{\xi}}_{s1}$ and $\vec{\vec{\xi}}_{s2}$ are
the
friction tensors of two isolated spheres; the indexes 1 and 2
referring to the two nearest neighbors.  In the limit when the
particle is far from the line, the friction tensor tends to the
well-known expression $\xi\vec{\vec{1}}$, with $\vec{\vec{1}}$ being
the unit matrix and $\xi=6 \pi\eta a$ with $\eta$ the viscosity of
the
fluid.  Therefore, both the tensorial character and the spatial
dependence of the friction are due to the presence of other objects.
In particular, due to stick boundary conditions, the friction tensor
diverges when the particle touches any solid surface, the divergence
being different depending on the direction of movement.  Lubrication
forces then hinder any contact between objects.  Explicit analytic
expressions for these tensors, which depend on the distance between
the particle and the line and on the relative position of the
particles, have been given by Brenner \cite{Bre} for the case of
sphere-plane friction (eqs.(2.19) and eqs.(2.63 and 2.65a)
respectively) and by Jeffrey-Onishi (eqs.  3.20,4.19,5.9 and 7.14
)\cite{JO} for the sphere-sphere friction tensor.  It should be noted
that, although particles are forced to adsorb on a line, the
expressions used for the friction tensors are three-dimensional, as
we
have real three-dimensional adsorbing spheres.

  In order to gain a physical understanding of the results obtained
in
the adsorption kinetics, it is interesting to study the effect of an
adsorbed particle on the dynamics of another one coming from the
bulk.
To this end, we have numerically solved eq.(\ref{eq1}).  In the
region
where the particles are close together, the mobility decays rapidly
to
zero.  Therefore, the numerical algorithm should contain a variable
time step in order to ensure that the mobility element associated
with
the perpendicular motion of the particles does not change
significantly during one integration step.  The unit of length has
been taken as the diameter of the spheres, and the unit of time $9
\nu
\rho_{f}/(a g \Delta\rho)$, with $\nu$ the kinematical viscosity of
the fluid.  In these units, eq.(\ref{eq1}) is dimensionless and,
therefore, does not depend on either the kind of particles or on the
medium.  This means, that the final position of the incoming
particles
will not depend on their mass or volume.  The initial conditions are
such that the incoming particle is at a heigth of $50$, at which HI
are negligible, and the horizontal position starts close to one of
the
adsorbed particles and is progressively displaced from the axis.  In
this way, we have studied the final position of the incoming sphere,
$x_{f}$, as a function of its initial distance to the axis, $x_{d}$
(see Fig.  1a).  Due to the divergence of the friction tensor when
the
particles are in contact, the calculation stops when the sphere is
almost touching the plane.  vdW are effectively taken into account by
stopping the simulation close to the plane.  When $d$ is large enough
, the results are plotted in Fig.  1b where we compare our results
with the predictions of BM, and from which we can infer the
modifications introduced by HI.  In the vicinity of the attached
particle, these corrections are to the order of 10-15\% with respect
to BM.  They originate from the fact that HI cause an increase of the
friction coefficient of the particle, which depends on its relative
motion with respect to the adsorbed one.  Thus, whereas the friction
coefficient for the perpendicular motion of two close objects at a
distance $r\equiv s+2 a$ diverges as $1/s$, the corresponding
coefficients for parallel motion diverge as $\ln s$.  The result is
an
effective repulsion between the particles.  We have found that the
final position predicted by HI differs from the one given by BM up to
values $x_{d}\sim5$.  Thus, the effects of HI persist far from the
attached particle.  For finite $d$ the discrepancies with respect to
BM are restricted to a closer region in the vicinity of the attached
particle but are essentially of the same order of magnitude.  This
effective repulsion will be responsible for the differences observed
in the adsorption process.

	We have numerically studied the deposition of spheres on a
line.  For this purpose, we have considered a line of length 800.  At
a height of 50, positions are chosen randomly.  Once one position has
been selected, it is taken as the initial condition for the sphere,
and subsequently eq.(\ref{eq1}) is solved numerically until the
sphere
reaches the line.  In order to speed up the program, if the incoming
particle nearly touches a preadsorbed one, and due to the results
shown in Fig.  1b., it is accepted in the line according to BM rules.
If room is available, the particle is adsorbed, otherwise it is
rejected.  Then, another position is randomly selected, and the
process goes on until either a prescribed fraction of covered line,
$\theta$, or the jamming configuration $\theta_{\infty}$, in which
there is no more room available for incoming particles, is reached.
Using periodic boundary conditions, we then generate configurations
of
adsorbed particles from which we can study the relevant quantities
related to this process.

	In this sense, we have analyzed the radial distribution
function, $g(r)$, of the adsorbed particles at different
concentrations.  In order to obtain representative values of $g(r)$,
approximately 1500 realizations of the adsorption process have been
performed.  In Fig.  2a the $g(r)$ at coverages $\theta=0.25$,
$\theta=0.5$ and at the jamming limit are displayed.  As expected,
$g(r)$ decays faster and the height of the peaks increases when
increasing the coverage, as in the BM case, although the initial
decay
after the peak does not depend too much on it.  The differences
between the $g(r)$ corresponding to their coverages are not
significant at distances larger than 3.  Furthermore, in Fig.  2b we
have compared $g(r)$ with the one given by BM.  We have observed a
smooth decay of the function behind the peak in contrast with the
sharp decay predicted by the BM model.  In the region of the first
peak large differences are observed, for example, at $r=1.3$ it is of
order 12\%, and of 42\% at $r=1.2$.  The $g(r)$ when HI are
considered
tends smoothly towards the corresponding quantity predicted by BM and
at $r=1.5$ the difference is 6\%.  These differences tend to decrease
when increasing the coverage, although even at jamming some
diferences
are observed.

	To investigate the effect of HI on global quantities, from
our
model we have computed the available fraction of line as a function
of
the coverage.  Figure 3 shows that differences, which are always
smaller than 1-2\%, are less significant than the ones obtained for
the radial distribution function.  Moreover, it has no relevant
effect
on the time evolution of the coverage either, as shown in the insert
of Fig.  3, whose assymptotic temporal behavior is equal to the one
prescribed by BM.  At intermediate times the coverage deviates from
BM
predictions, leading to a jamming limit $\theta_{\infty}=.797$,
slightly smaller than the BM one, $\theta_{\infty}^{BM}=.810$.

    In summary, we have shown that in respect to macroscopic
quantities of the deposition process, our model introduces no
important corrections to BM results.  The explanation of this result
lies in the fact that these quantities can be considered as averages
over the line, and therefore local details are masked.  However,
local
properties are strongly affected by HI.  In particular, the
distribution of spheres around a preadsorbed one at low coverages
shows differences up to 10\% due to the effective repulsion induced
by
HI.  In regards to the radial distribution function, the decay after
the peaks is slower than in BM.  Behind the first peak, differences
can be as large as 40\%, and the convergence towards BM is slow.
Discrepancies after the second peak are also observed, though they
are
smaller.  This behavior decreases slightly with the coverage, so that
even at jamming some differences are observed.  Our conclusion is
that
HI cannot be neglected a priori when studying such properties.
Recent
experimental results for the adsorption of colloidal particles on a
surface\cite{Sch1} show that the radial distribution function
deviates
from BM predictions, which is the same kind of behavior shown in Fig.
3.  Though our simulations have been perfomed in 1d, they capture the
essential features of the process, and explain the differences
between
experiments and BM predictions, which do not originate from the
polydispersity of the solution \cite{Sch2}.  In 2d these differences
can be expected to be smaller than the ones reported in this Letter
because of the additional angular average performed to arrive at the
radial distribution function.  BM has been introduced as the limiting
case when the deposition is controlled by gravity instead of
diffusion\cite{Tal}.  However, we have shown that in the regime when
gravitational effects become important, HI introduce significant
effects.  BM could then reasonably describe the physics of the
deposition process only when inertial effects become dominant, since
in this regime the damping term is negligible.  This fact occurs for
times not larger than the inertial time $\tau\equiv m/\xi=2
a^{2}\Delta\rho/(9 \nu \rho_{f})$.  In usual experimental situations,
as the one reported in ref.\cite{Sch1}, $\tau\sim 10^{-6}s$, while
the
experimental time scale is of the order of minutes.  Consequently, in
this situation inertial effects are negligible, which explains the
disagreement between the radial distribution function obtained
experimentally and the one calculated from BM, and justifies the
validity of eq.(\ref{eq1}).  Therefore, the applicabilitty of BM is
severely restricted in experimental situations.  Finally, the fact
that HI affect the local distribution of adsorbed particles implies
that these interactions will be relevant when studying other physical
properties of adsorbed layers, as for example the dielectric
susceptibility of adsorbed particles \cite{Dick}.

\acknowledgements

	We would like to acknowledge Profs.  P.  Schaaf and D.
Bedeaux for fruitful discussions.  This work has been supported by
the
European Economic Community under grant SCI$^{*}$-CT91-0696 and by
CICYT (Spain), grant PB92-0895.

\begin{figure*}

\caption{a) Illustration of the geometry of the model.  An incoming
particle is selected at a height $h_{0}$ at a certain distance
$x_{d}$ from the nearest particle.  A second nearest neighbor is
located at a distance $d$.  The particle will end in the position
$x_{f}$.  b) Final position as a function of the initial displacement
when $d$ is large enough.  Insert, the difference in the final
position predicted by HI and BM relative to the one given by BM.}

\caption{a) g(r) for the HI model at coverages $\theta=0.25, 0.5$
and jamming. b) Comparison with BM at $\theta=0.5$.}

\caption{ Available line fraction as a function of the coverage
$\theta$ for BM and with HI.  Both curves are practically
indistinguishable up to high coverages.  Insert, the coverage as a
function of time for BM and HI.}

\end{figure*}

\end{document}